\documentclass[11pt,letter]{article}

\usepackage{poma_style}
\usepackage{multicol}
\usepackage{graphicx}
\usepackage{amsmath,amsfonts,bm}
\usepackage{courier}
\usepackage[colorlinks=true,urlcolor=blue,linkcolor=black,citecolor=black]{hyperref}
\usepackage[utf8]{inputenc}
\usepackage{doi}
\usepackage[T1]{fontenc}

\author{
  Ziemer, Tim\\
  Bremen Spatial Cognition Center, University of Bremen\\
  \texttt{ziemer@uni-bremen.de}
    \and
  Kiattipadungkul, Pattararat\\
  Faculty of Information and Communication Technology, Mahidol University\\
  \texttt{pattararat.k@gmail.com}
   \and
   Karuchit, Tanyarin\\
   Faculty of Information and Communication Technology, Mahidol University\\
     \texttt{tanyarin.kar@student.mahidol.ac.th}
     }
\date{}

\title{Novel Recording Studio Features for Music Information Retrieval}

\begin{document}
\maketitle
\abstract{Producers of Electronic Dance Music (EDM) typically spend more time creating, shaping, mixing and mastering sounds, than with aspects of composition and arrangement. They analyze the sound by close listening and by leveraging audio metering and audio analysis tools, until they successfully created the desired sound aesthetics. DJs of EDM tend to play sets of songs that meet their sound ideal. We therefore suggest using audio metering and monitoring tools from the recording studio to analyze EDM, instead of relying on conventional low-level audio features. We test our novel set of features by a simple classification task. We attribute songs to DJs who would play the specific song. This new set of features and the focus on DJ sets is targeted at EDM as it takes the producer and DJ culture into account. With simple dimensionality reduction and machine learning these features enable us to attribute a song to a DJ with an accuracy of 63\%. The features from the audio metering and monitoring tools in the recording studio could serve for many applications in Music Information Retrieval, such as genre, style and era classification and music recommendation for both DJs and consumers of electronic dance music.}

\section{INTRODUCTION}
Electronic Dane Music (EDM) is a genre of music characterized as having a repetitive beat and synthesized backing audio \cite{djc}. EDM is an umbrella term, also referred to as electronica, techno or electro in their widest sense\footnote{An overview can be found on \url{http://techno.org/electronic-music-guide/}. Note that we use the term without any implications concerning commercial intentions, level of originality, or artistic value.}. EDM is produced primarily using electronic instruments and computers and is comprised of many subgenres and styles. EDM is often created and performed by Disc Jockeys (DJs) \cite{djcult}. They generally select and mix EDM songs for broadcast, live performances in night clubs, concerts and on music festivals. Similarly to musical subgenres, the music that is curated by a DJ turns into a category of its own. The AllMusic Guide \cite{allmusic} describes ``Electronic'' music as ``\ldots a relentless desire to find a new sound no matter how tepid the results." \cite[p. 107]{cult} puts it: ``DJs such as Armin van Buuren, Paul van Dyk, Paul Oakenfold, Sasha and Ti{\"e}sto pushed EDM into mainstream consciousness, [and] expanded the possibilities of the electronic dance soundscape (\ldots) the superstar DJ has become (\ldots) a significant arbiter of musical taste.'' Consequently, many DJs stick to a certain sound rather than (sub-)genre. In practice, music producers achieve the desired sound by a combination of close listening and audio metering and monitoring  tools in the recording studio. Furthermore, DJs contribute in steering, shaping or even defining musical taste of many music listeners.

In this paper we suggest a set of features derived from typical audio metering and analysis tools in the music studio. These features should indicate a causal relationship between the objective acoustics and the subjective sound preference of EMD producers and consumers. First, we analyze the music that specific DJs integrated in their sets by means of the proposed features. Then, we train a machine learning algorithm to estimate which music piece matches the sound ideal of which DJ, i.e., to predict which DJ would play which song. In that sense, DJs are the classes, rather then genre or mood of a song. The two facts that 1.) a DJ's job is to select appropriate music for a large audience, and 2.) every DJ has his/her own sound ideal, are a promising starting point for music classification that can serve, e.g., for automatic music recommendation for EDM \cite{pim}. If a computer could identify a DJ's sound ideal, it could analyze millions of songs and recommend those songs that seem to meet this sound ideal. Such a music recommendation approach could for example serve to enlarge the repertoire of a DJ by recommending him/her matching but yet unknown music. Or it could serve as a music recommendation tool for music consumers in music stores and streaming services. They would simply specify their favorite DJ and the trained machine learning algorithm would recommend music that seems to meet this DJ's sound ideal. It could even serve the other wary round; as a DJ recommendation tool: A user lists some seed songs and the machine could recommend him/her a DJ whose sound ideal matches the seed songs best.


The remainder of this paper is structured as follows. Section \ref{related} gives a broad overview about audio content-based music classification and recommendation. Then, the music dataset we used is described in detail in Section \ref{dataset}. Section \ref{method} describes our method. It is subdivided into one subsection that explains the selection of DJs, segmentation of music pieces and feature extraction. The second subsection describes the dimensionality reduction and machine learning approach. The classification results are presented and discussed in Section \ref{results}, followed by a conclusion in Section \ref{conclusion} and an outlook in Section \ref{outlook}.

\section{RELATED WORK}
\label{related}
Nowadays, online music stores and streaming services provide access to millions of songs. Both music providers and customers need tools that organize the music and help the customer exploring unknown songs finding music of their personal taste. Therefore, two topics are of relevance in Music Information Retrieval (MIR) research; music classification and recommendation, which are briefly introduced in this section based on some exemplary cases.

\subsection{Acoustically-Based Genre Classification}
Automatic genre classification by means of acoustic features dates back to the early years of Music Information Retrieval. One early example is \cite{tzanetakis}, who extract about $10$ features and test a number of classifiers to classify music into $190$ different genres with an accuracy of $61$\%.  A large number of papers addressed this problem, mostly focusing on improved redundancy reduction, feature selection and kernel optimization for the classifier. One example is  \cite{later} who defined an elaborated kernel function for a support vector machine (SVM) and test multiple feature selection and dimension reduction methods to achieve a genre classification accuracy of $89.9$\% .

However, \cite{horse} convicted most of these approaches as being a \emph{horse}, i.e., classifying not based on causal relationships between acoustic feature and genre, but based on statistical coincidences inherent in the given dataset. This means, that many of the commonly extracted low-level features are meaningless for the task of genre classification, and the classifier is not universally valid, i.e., it is not likely to perform similarly well when applied to another dataset. Most importantly, a large number of publications showed that sophisticated and elaborate machine learning approaches paired with extensive model testing can tune classification accuracy from $60$\% to about $90$\%, even when the features are not explanatory for the classes. This fact can lead to the erroneous assumption that an accurate classifier is a reliable and/or explanatory classifier.

Another weakness of automated genre classification is the lack of a valid ground truth. Genre and style descriptions are debatable and may depend on aspects, such as decade, music scene, geographic region, age, and many more \cite{classw}. In other words, genre, subgenres and styles are typologies, rather than classes. In contrast to classes, typologies are neither exhaustive nor mutually exclusive, not sufficiently parsimonious,  based on arbitrary and ad hoc criteria and descriptive rather than explanatory or predictive\cite{typologies}. A questionable ground truth is used to train a model. Consequently, the model output is at best questionable, too.

In contrast to genre labels, DJ sets are a documented ground truth. However, it must be mentioned that DJs are not always free to play exclusively music that matches their sound ideal, but also music that is requested, e.g., by the audience or their record company. It is known that the nationality of DJs plays a certain role of influence on their performance style \cite[p. 3 and 210]{bobby}. Furthermore, DJs need to play new hit records and pre-released songs to stay exclusive and up-to-date. These circumstances may enforce them to play music that does not fully meet their aesthetic sound ideal. In addition to that, the taste and sound ideal of DJs is nothing static but may change over time.



\subsection{Acoustically-Based Music Recommendation}
Social services like Last.FM\footnote{See \url{https://www.last.fm}.} capture the music listening behavior of users and let them tag artists and songs from their local media library, which can serve as a basis for collaborative filtering based on tags and/or listening behavior. A common issues with collaborative filtering is the so-called cold start problem \cite{deng}, i.e., new songs have only few listeners and no tags. This can be avoided by content-based music recommendation, e.g., based on metadata, such as artist, album, genre, country, music era, record label, etc. This approach is straight-forward, as these information are largely available in online music databases, like discogs\footnote{See \url{https://www.discogs.com}.}. Services like musicube\footnote{See \url{http://musicu.be}.}  scrape information --- e.g., about producer, songwriter, mood and instrumentation --- to let end users of online music stores and streaming services discover a variety of music for all possible key terms and phrases. However, issue with recommendation based on music databases are the inconsistency of provided metadata and the subjectivity of some tags, especially mood- and genre-related tags. Acoustically-based music recommendation can avoid these issues, as they are based on acoustic features that can be extracted from any audio file.

\cite{deng} let users rate the arousal, valence and resonance of music excerpts as a ground truth of music emotion. Then, they extract dozens of acoustic features and associate them with either of the dimensions. They use a dimensionality-reduction method to represent each dimension based on the associated features with minimum squared error. Finally, based on an analyzed seed song, the recommender suggests music pieces from a similar location within that three-dimensional space.

Likewise, the MAGIX mufin music player analyzes a local music library and places every music piece at one specific location within a three-dimensional space based on audio features, see, e.g., \cite[chap. 2]{buch}. Here, the three dimensions are labeled synthetic/acoustic, calm/aggressive, and happy/sad. The software recommends music from an online library that lies very near to a played seed song. The semantic space certainly makes sense, as it characterizes music in a rather intuitive and meaningful way. Drawbacks of the approach are that 1.) listeners may disagree with some of the rather subjective labels, 2.) the transfer from extracted low-level features to semantic high-level features is not straight-forward and it can be observed that the leveraged algorithm often fails to allocate a music piece correctly, at least along the rather objective synthetic/acoustic-dimension.

\cite{womrad} extract over $60$ common low-level features to characterize each song from a music library. Users create a seed playlist that they like. Then songs with a similar characteristic are recommended. The authors test different dimensionality reduction methods and distance models. In the most successful approach users rated $37.9$\% of the recommendations as good. Compared to that, recommending random songs with the same genre tag as most of the songs in the seed playlist created no less than $66.6$\% good recommendations. Still, the authors argue that their approach creates more interesting, unexpected and unfamiliar recommendations. This is desired because users need novel inspiration instead of obvious choices from a recommender.

\cite{fire} aim to identify inter-subjective characteristics of music pieces leveraging psychoacoustic models of loudness, roughness, sharpness, tonalness and spaciousness. A seed playlist is analyzed and then songs with similar psychoacoustic magnitudes are recommended. No machine learning is applied. Instead, the psychoacoustic features are assumed as being orthogonal and a simple Euclidean distance is calculated. In a listening test, users rated $56$\% of the recommendations as good.

It can be observed that all these music recommendation approaches focus on the music consumer. Due to our features and the focus on DJs of Electronic Dance Music, our classifier could serve as a basis for a music recommendation tool not only for music consumers but also for DJs.

\section{THE DATASET}
\label{dataset}
For the sake of this study a small EDM database was created and analyzed.

\subsection{Data acquisition}
When a disc jockey performs, he or she plays many songs in a single set. The scale of our dataset is as follows: in total, the $10$ most popular DJs according to DJMag \footnote{See \url{https://djmag.com/top100dj}, retrieved June 4 2019.} were selected, each with $10$ DJ sets. Each DJ set is about one to two hours long and consists of multiple tracks. The tracklist of each DJ set can be found on the 1001Tracklists website\footnote{See \url{https://www.1001tracklists.com}, retrieved June 4 2019.}. As a result, a total of $1,841$ songs, often in a specific version or remix, were identified from $100$ DJ sets. The artist names of the $10$ DJs are:
\begin{enumerate}
\item Martin Garrix
\item Dimitri Vegas \& Like Mike
\item Hardwell
\item Armin van Buuren 
\item David Guetta
\item Dj Ti{\"e}sto
\item Don Diablo
\item Afrojack
\item Oliver Heldens
\item Marshmello
\end{enumerate}
\subsection{Data cleaning}
We purchased all songs in the respective version or remix that we could find in online music stores; $1,262$ songs in total. In a first step, the compressed audio files were decoded to PCM files with a sample rate of $44,100$ Hz, a sample depth of $16$ bit and $2$ channels. The file was cropped to remove silence in the beginning and at the end of the file, and then normalized to $0$ dB, i.e., 

\begin{equation}
\text{Max}\left[\left|x_m\right|\right]=1 \ ,
\end{equation}
where $x_m$ represents the $m$th sample of the cropped PCM file, and can take values between $-1$ and $1$. Many club versions of songs begin and end with a rather stationary drum loop that makes it easier for a DJ to estimate the tempo, to blend multiple songs and to fade from one song to the next. We consider these parts as a practical offer to the DJ rather than a part of the creative sound design process. Consequently, we eliminate it by analyzing only the central $3$ minutes of each song.

The audio files were divided into segments of $N = 2^{12} = 4,096$ samples, which corresponds to about $93$ ms. From each segment, a number of features was extracted. The features represent audio metering and monitoring tools that are commonly used in recording studios when mixing and mastering the music. They are introduced in the following section.
\section{METHOD}
\label{method}
We analyze EDM through features that describe what audio metering and monitoring tools in the recording studio display. Producers of EDM use the same tools to create their desired sound. DJs make a selection of songs with matching sounds.

\subsection{Features}
Audio monitoring tools help music producers and audio engineers to shape the sound during music mixing and mastering\cite{secrets,protools,meter,current}.  For producers of EDM, the sound plays an exceptionally large role. They listen closely to the mix and consult such tools to achieve the desired sound in terms of temporal, spectral, and spatial aspects. Some audio monitoring tools are used on the complete stereo signal, whereas others are also applied to each third-octave band.

\subsubsection{Volume Unit}
\begin{figure}
   \begin{minipage}[t]{.4\linewidth} 
	\includegraphics[width=2.45in]{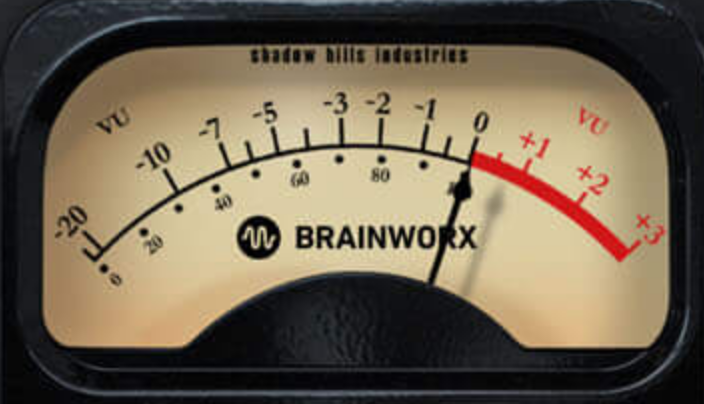}
	\caption{\label{fig:vu}A screenshot of the VU meter in the brainworx Shadow Hills Mastering Compressor plugIn.}
   \end{minipage}
   \hspace{.1\linewidth}
   \begin{minipage}[t]{.4\linewidth} 
	\includegraphics[width=2.45in]{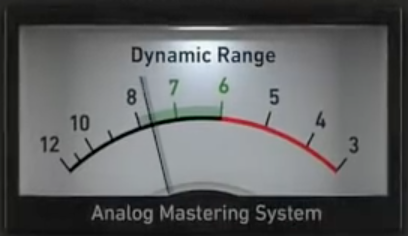}
	\caption{\label{fig:dr}A screenshot of the dynamic range meter in the Brainworx  bx\textunderscore masterdesk plugIn.}
   \end{minipage}
\end{figure}

The Volume Unit (VU) meter \cite[chap. 5]{sr}, \cite[chap. 7]{protools} is a volt meter, indicating an average volume level that had been standardized for broadcast between the 1940s and 1990s \cite[chap. 12]{meter}. Originally, an electric circuit deflected a mechanical needle. The inertia created a lag smearing that loosely approximated the signal volume over an integration time of $300$ ms. In digital recording studios, a pseudo VU meter can be implemented by calculating
\begin{equation}
\text{VU} = 20 \log_{10}\left(\frac{\sum_{d=1}^{D}{\left|x_d\right|}}{\sum_{d=1}^{D}{\left|\sin(2\pi t 1000 \text{Hz})\right|}}\right) \ .
\end{equation}
Here, the denominator is a $1$ kHz tone with maximum amplitude. Originally, $D$ should be $13,230$ samples at a sample rate of $44,100$ Hz, which corresponds to a duration of $300$ ms. An example is illustrated in Fig. \ref{fig:vu}. In our case $D=N$. 

\subsubsection{Peak Programme Meter}
\begin{figure}[t]
 \begin{minipage}[b]{.28\linewidth}
	\includegraphics[width=1.7in]{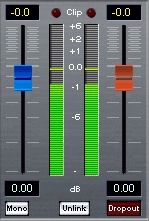}
	\caption{\label{fig:ppm}A screenshot of the Steinberg Wavelab bargraph meter indicating the current level as a bar, the short-term peak level as a single line above the bar, the overall peak as a number in yellow, and it indicated cipping by red LEDs.}
\end{minipage}
  ~~~~~  
 \begin{minipage}[t]{.3\linewidth}
	\includegraphics[width=1.7in]{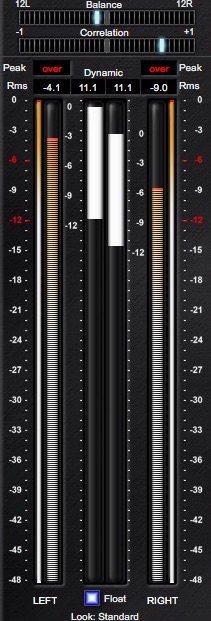}
	\caption{\label{fig:bw}A screenshot of the brainworx bx\textunderscore meter plugIn, which indicates signal RMS, peak, DR, channel correlation and balance.}
\end{minipage}
  ~~~~~  
 \begin{minipage}[b]{.3\linewidth}
	\centering
	\includegraphics[width=1.7in]{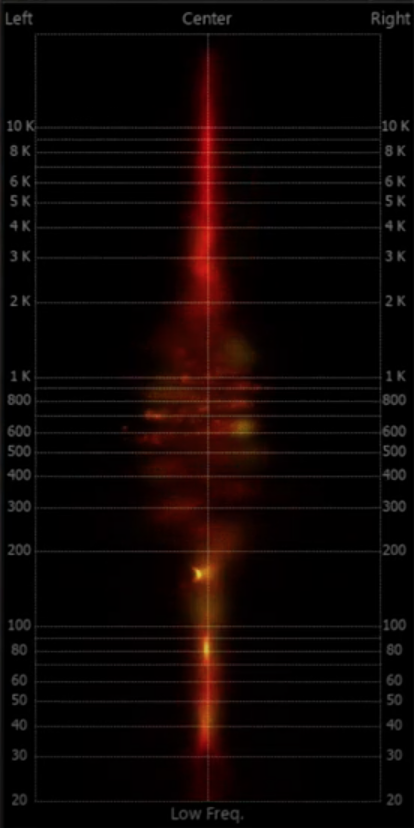}
	\caption{\label{fig:pan2}A screenshot of the FLUX:: Pure Analyzer plugIn. Here, the pan of each frequency band is indicated as a horizontal deflection.}
\end{minipage}
\end{figure}

The peak meter is the standard level meter \cite[chap. 7]{protools}. The Peak Programme Meter (PPM) resembles a VU meter but has a much shorter integration time. In our digital case the PPM meter gives the peak value of a time frame
\begin{equation}
\text{PPM} = 20 \log_{10}\left(\text{Max}\left[\left|x_n\right|\right]\right) \ .
\end{equation}
In the recording studio these are displayed by bargraph meters\cite[chap. 5]{sr}, as can be seen in Fig. \ref{fig:ppm}.

\subsubsection{Dynamic Range}
The Dynamic Range (DR) indicator \cite[chap. 3]{secrets} expresses the dynamic range within one time segment as ratio between lowest and largest peak within sub-segments $s_i$ of $10$ ms duration as
\begin{equation}
\text{DR} = 20 \log_{10} \left(\frac{\text{Max}\left[s_i\right]}{\text{Min}\left[s_i\right]} \right) \ .
\end{equation}
One DR meter is illustrated in Fig. \ref{fig:dr}. It is related to the crest factor, also referred to as peak-to-average ratio \cite[p. 293]{protools}. Both are low for mellow instruments and rise with increasing percussiveness \cite{Schneider2018}.

The above-mentioned audio metering tools are typically applied on single tracks and the complete master output. The audio monitoring tools in the following sections are applied to the complete audio as well as to each individual third-octave band.

\subsubsection{Third-Octave Bands}
In addition to the complete audio, the following audio monitoring tools are also applied on each of the $27$ third-octave bands around the center frequencies $40$, $50$, $63$, $80$, $100$, $125$, $160$, $200$, $250$, $315$, $400$, $500$, $630$, $800$, $1000$, $1250$, $1600$, $2000$, $2500$, $3150$, $4000$, $5000$, $6300$, $8000$, $10000$, $12500$, and $16000$ Hz \cite[chap. 13]{sr} filtered according to the ANSI s1.11 standard\cite{ansi}.

\subsubsection{Root Mean Square}
In the recording studio the Root Mean Square (RMS) of the audio time series is used to approximate loudness\cite[chap. 5]{meter}, \cite[pp. 301ff]{protools}. The RMS is the quadratic mean calculated as
\begin{equation}
\text{RMS} = \sqrt{\frac{\sum_{d=1}^{D}{x_d^2}}{D}} \ , 
\end{equation}
where $x_d$ represents the $d$th sample in the PCM file. Audio monitoring tools, tend to indicate the RMS in combination with other audio meters, like DR, peak, channel correlation and channel balance, as illustrated in Fig. \ref{fig:bw}.

\begin{figure}
	\centering
	\includegraphics[width=4in]{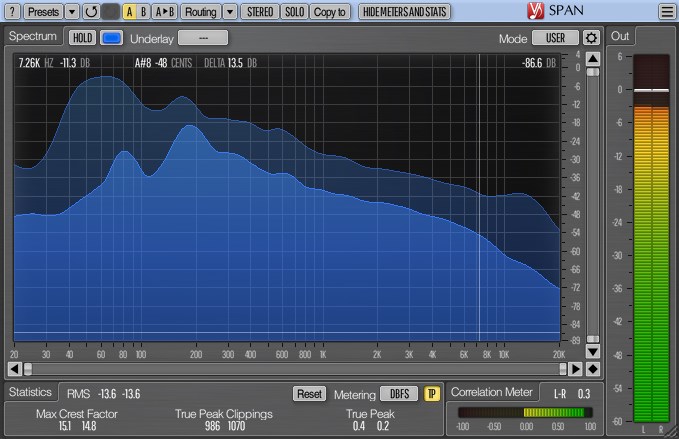}
	\caption{\label{fig:spec}A screenshot of the Voxengo SPAN plugIn, which indicates the peak (dark blue) and RMS (light blue) of each third octave band, together with the full bandwidth peak, RMS, crest factor and channel correlation.}
\end{figure}

\subsubsection{Phase Scope Distribution}
The phase scopes, also referred to as vector scope or goniometer, simply plot the left over the right channel, tilted by $45^\circ$ \cite[chap. 18]{meter}\cite{mores}, \cite[chap. 7]{protools}. Music producers derive a lot of information from qualitative and quantitative phase scope analysis.

In the recording studio, a distribution along a vertical line indicates positive channel correlation and thus mono compatibility. Vertical lines are perceived as narrow, clearly localizable sound sources. In contrast, a distribution along a horizontal line indicates negative channel correlation, which creates destructive interference in the stereo loudspeaker setup. The optimal mixture of clear and diffuse sound is an important aspect of the producers sound ideal and is achieved through close listening and qualitative analysis of the phase scope distribution.

\begin{figure}
\begin{minipage}[t]{.25\linewidth}
	\includegraphics[width=1.7in]{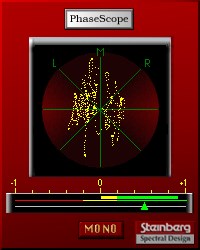}
	\caption{\label{fig:me}A screenshot of the Steinberg PhaseScope plugin, which plots the left over the right channel for each time window (yellow distribution on top) and indicates the current, highest and lowest channel correlation (horizontal bar at the bottom).}
\end{minipage}
 ~~
\begin{minipage}[t]{.25\linewidth}
	\includegraphics[width=1.7in]{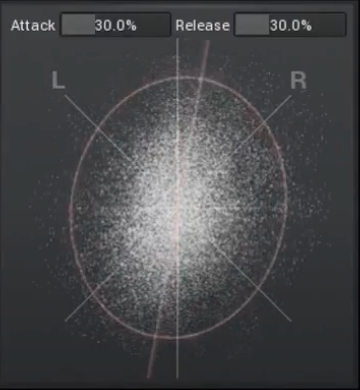}
	\caption{\label{fig:pan}A screenshot of the MeldaProduction MStereoProcessor plugIn. This phase scope meter indicates the stereo width of a complete mix with an ellipse and the pan is given by its major diameter.}
\end{minipage}
 ~~
\begin{minipage}[t]{.4\linewidth}
	\includegraphics[width=2.6in]{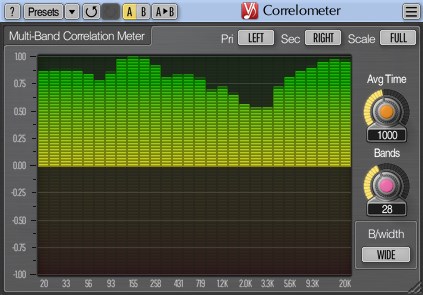}
	\caption{\label{fig:vox}A screenshot of the Voxengo Correlometer plugIn, which indicates the channel correlation for a variable time segment length in each third octave band.}
\end{minipage}
\end{figure}

We imitate this qualitative analysis by using box counting as suggested in \cite{klg}. First, we divide the two-dimensional space into $20$ times $20$ boxes. Then, we count how many of the $400$ boxes are occupied by one or more values. Figure \ref{fig:me} illustrates a phase scope plugin.

\subsubsection{Phase Scope Panning}
A second feature that is derived from the phase scope is the panning, i.e., the dominant angle in the distribution. We calculate it by plotting the absolute value of the left against the absolute value of the right channel and transforming the coordinates from Cartesian to polar coordinates. Then, we calculate the mean angle. Typically, dominant kickdrums tend to attract the distribution near $0^{\circ}$, whereas reverberant pads, strings and atmo sounds tend to create a somewhat random distribution. Figures \ref{fig:pan} and \ref{fig:pan2} show two audio monitoring tools that indicate the pan of single frequency bands and the complete signal.

\subsubsection{Channel Correlation}
Another measure often implemented in phase scope tools is Pearson's correlation coefficient $P$ of the two stereo channels \cite{klg,current}. It is sometimes referred to as \emph{phase meter} or \emph{correlation meter} \cite[chap. 7]{protools}. It can be calculated as
\begin{equation}
P = \frac{\text{cov}\left(L,R\right)}{\sigma\left(L\right) \sigma\left(R\right)} \ ,
\end{equation}
where $L$ and $R$ represent the left and right channel vectors, $\text{cov}$ the covariance and $\sigma$ the standard deviation, or as 

\begin{equation}
P = \text{Max}\left[\left|L\ast R\right|\right] \,
\end{equation}
where the asterisk denoted the cross-correlation operation. Here high positive values indicate mono compatibility, and a comparably narrow sound impression. Negative values indicate phase issues and violation of mono compatibility.

Many producers like to keep the overall channel correlation near $0.1$, whereas producers that like prominent drum sounds may prefer much higher values, and producers of atmospheric sounds may even aim at negative channel correlation. A channel correlation meter for third-octave bands is illustrated in Fig. \ref{fig:vox}.

\subsection{Dimensionality Reduction}
Together, the above-mentioned audio metering and monitoring tools yield a $146$-dimensional feature vector:
\begin{multicols}{3}
\begin{itemize}
\item $8$ features à $2$ channels:
\subitem VU
\subitem PPM
\subitem DR
\subitem RMS
\item[\vspace{\fill}]
\item $3$ features on stereo:
\subitem Box counting
\subitem Panning
\subitem Channel correlation
\item[\vspace{\fill}]
\item[\vspace{\fill}]
\item $5$ features à $27$ bands:
\subitem RMS (left)
\subitem RMS (right)
\subitem Box counting
\subitem Panning
\subitem Channel correlation
\end{itemize}
\end{multicols}
Each feature contains $1,983$ values: one value for each window. These features are not orthogonal, but somewhat redundant. 

A Principal Component Analysis (PCA) is applied to represent the the $146$-dimensional space by the one diagonal line that that minimizes the squared distance between the line and all data points, i.e., the first principal component.

A PCA is an established, linear method for dimensionality reduction based on feature projection. More sophisticated, non-linear dimensionality reduction approaches, such as self-organizing maps\cite{som}, are out of scope of this paper, but may improve the information content, and thus the success of the classification, a lot.

\subsection{Classification}
To test the usefulness of audio metering and monitoring tools as features for tasks related to EDM, we use the tools to classify $10$ different DJs. We use random forest model \cite{random}, which is made up of hundreds or thousands of decision trees. Only a subset of all the features are considered for splitting each node in each decision tree. The final predictions are made by averaging the predictions of each individual tree. Model fitting was done using sklearn\footnote{See \url{https://scikit-learn.org/stable/}.}. For the classifier we applied $5$-fold cross-validation, i.e., the original sample is randomly partitioned into $5$ equal sized subsamples.


\subsection{Test and Evaluation}
Cross validation is used to evaluate the classification models. The dataset is then split from the dataset with the \protect\path{train_test_split()} function from the Python scikit-learn machine learning library to separate the data into a training dataset and a testing dataset. $90$\% of the dataset was used for training while the remaining $10$\% of the data was used for testing. The performance of classification predictive model is evaluated using an accuracy score. 

\section{RESULTS AND DISCUSSION}
\label{results}
The random forest algorithm returns an overal accuracy of $62.99$\%. The result shows that the highest accuracy is achieved when \protect\path{n_estimators} is $25$, bootstrap is \emph{True} and the criterion for classification is set to \emph{entropy}. Other parameters include the maximum depth of the tree of $15$, the random state of $49$, and the number of features to consider when looking for the best split is \emph{auto}. Note that for \emph{auto}, $\text{max}\left[\text{features}\right] = \sqrt{n_\text{features}}$.

Table \ref{tab:report} shows the confusion matrix of the DJs. In $8$ out of $10$ cases the most frequent model output is the correct DJ. The two exceptions are Armin van Buuren and Hardwell. Armin van Buuren has frequently been falsely labeled as Dj Ti{\"e}sto in $5$ out of $11$ cases and Oliver Heldens as Don Diablo in $3$ out of $11$ cases. These confusions are not necessarily a weakness of the features or the machine learning. Both artist pairs appear to be quite similar. Last.FM lists Dj Ti{\"e}sto as the most similar Dj to Armin van Buuren and Don Diablo as most similar to Oliver Heldens\footnote{See \url{https://www.last.fm/music/Tiesto/+similar} and \url{https://www.last.fm/music/Oliver+Heldens/+similar} -- retrieved January 15 2021. On Last.FM artist similarity is based on user-generated tags \cite{lastfm}.}. All three Hardwell examples have been attributed to other DJs, which could be a result of the low number of Hardwell pieces in the test data.

\begin{table}[]
\begin{tabular}{ccccccccccc|ccc}
\textbf{Predicted/Actual DJ} & \rotatebox{90}{\textbf{Martin Garrix}} & \rotatebox{90}{\textbf{D. Vegas \& L. Mike}} & \rotatebox{90}{\textbf{Hardwell}} & \rotatebox{90}{\textbf{Armin van Buuren}} & \rotatebox{90}{\textbf{David Guetta}} & \rotatebox{90}{\textbf{Dj Ti{\"e}sto}} & \rotatebox{90}{\textbf{Don Diablo}} & \rotatebox{90}{\textbf{Afrojack}} & \rotatebox{90}{\textbf{Oliver Heldens}} & \rotatebox{90}{\textbf{Marshmello}}&\rotatebox{90}{\textbf{precision}}&\rotatebox{90}{\textbf{recall}}&\rotatebox{90}{\textbf{f1-score}}\\
\textbf{Martin Garrix}                   & $\mathbf{10}$         & $0$          & $0$          & $1$          & $2$          & $3$          & $0$          & $0$          & $0$          & $0$        & $0.833$ & $0.625$ & $0.714$ \\
\textbf{D. Vegas \& L. Mike}                   & $0$          &$\mathbf{6}$          & $0$          & $1$          & $0$          & $0$          & $0$          & $1$          & $1$          & $1$       & $0.667$ & $0.600$ & $0.632$   \\
\textbf{Hardwell}                   & $0$          & $\mathbf{1}$          & $0$          & $\mathbf{1}$          & $0$          & $\mathbf{1}$          & $0$          & $0$          & $0$          & $0$       & $0$ & $0$ & $0$  \\
\textbf{Armin van Buuren}                   & $0$          & $2$          & $1$          & $2$          & $0$          & $\mathbf{5}$          & $0$          & $0$          & $0$          & $1$     & $0.333$ & $0.182$ & $0.235$     \\
\textbf{David Guetta}                   & $1$          & $0$          & $0$          & $0$          & $\mathbf{5}$          & $0$          & $2$          & $2$          & $0$          & $0$     & $0.556$ & $0.500$ & $0.526$       \\
\textbf{Dj Ti{\"e}sto}                   & $1$          & $0$         & $1$          & $0$          & $0$          & $\mathbf{12}$         & $2$          & $0$          & $0$          & $0$   & $0.480$ & $0.750$ & $585$  \\
\textbf{Don Diablo}                   & $0$          & $0$          & $1$          & $1$          & $0$          & $2$          & $\mathbf{19}$         & $0$          & $1$          & $1$    & $0.704$ & $760$ & $731$         \\
\textbf{Afrojack}                   & $0$          & $0$          & $1$          & $0$          & $0$          & $1$          & $1$          & $\mathbf{2}$          & $0$          & $0$     & $0.333$ & $0.400$ & $364$       \\
\textbf{Oliver Heldens}                   & $0$          & $0$          & $0$          & $0$          & $1$          & $0$          & $3$          & $1$          & $\mathbf{6}$          & $0$    & $0.667$ & $0.545$ & $0.600$      \\
\textbf{Marshmello}                  & $0$          & $0$          & $0$          & $1$          & $1$          & $0$          & $0$          & $0$          & $1$          &$\mathbf{17}$    & $0.850$ & $0.850$ & $0.850$ \\
\hline
\textbf{support/avg.}&$16$&$10$&$3$&$11$&$11$&$16$&$25$&$5$&$11$&$20$&$0.634$&$0.622$&$0.619$
\end{tabular}
\caption{Confusion matrix of the random forest model.}
\label{tab:report}
\end{table}

The Table  also shows the classification report of the random forest model. \emph{Precision} is the ratio $y_\text{correct}/\text{freq}_y$. It indicates how many musical pieces have been attributed to the specific DJ correctly, divided by the total number of pieces that have been attributed to that DJ. This equals the number of correct classifications (diagonal cells) divided by the sum of the respective column. \emph{Recall} is the ratio $y_\text{correct}/\text{freq}_x$. It describes how many musical pieces have been attributed to the specific DJ correctly, divided by the total number of pieces from that specific DJ.  This equals the number of correct classifications (diagonal cells) divided by the sum of the respective row. Except for Dj Ti{\"e}sto the precision and recall scores of each DJ have a similar magnitude. Dj Ti{\"e}sto has a good recall, but, as explained above, the precision is somewhat lower because many songs from the  Armin van Buuren sets have been falsely attributed to him. \emph{Support} means how many songs of the respective DJ's sets occurred in the evaluation dataset. It is evident that Hardwell and Afrojack are poorly classified due to the low number of occurrences in the test dataset. The \emph{avg.} indicates the average precision, support and f1-score values.


\section{CONCLUSION}
\label{conclusion}
We proposed features derived from audio metering and monitoring tools commonly used in music recording studios.  In contrast to many conventional low-level features or psychoacoustically-motivated features, these features are consulted by music producers and recording engineers in practice. We hypothesize that such features can be valuable for music analysis, classification and recommendation, particularly in the field of electronic dance music, where sound plays a crucial role on a par with (or even more important than) other aspects such as composition, arrangement and mood. We evaluate these features by a classification task. An audio content-based classifier attributes songs to DJs who would play the song. The success rate of $63$\% and the fact that classification errors can be explained by artist similarity and the unbalanced data set are evidence that the proposed features give causal explanations of a DJs' sound ideal, which can be valuable for various music information retrieval tasks, such as sound analysis, genre classification and music recommendation, and maybe even producer recognition.

\section{OUTLOOK}
\label{outlook}
The proposed set of features seems suitable to give some explanation of a DJ's sound ideal from the viewpoint of recording studio technology and practice. Combining the audio metering and monitoring features with psychoacoustic features, which attempt to mimic the producer's auditory perception during close listening, might be the perfect match to achieve a deep understanding of music production aesthetics, sound preference and musical taste. Naturally, it would be beneficial to validate the features by means of a listening experiment, either with the respective DJs, or with listeners of electronic dance music.

According to many music recording, mixing, and mastering engineers, sound plays an important role for all types of music, not only for EDM \cite[p. 400]{manual}, \cite[p. 240]{roads}, \cite[p. V]{gibson}. It may be interesting to see how the proposed features perform on other popular music, like pop, rock, hip-hop, jazz, and on recordings of classical music, raaga, gamelan or sea shanty pieces.
\section*{Acknowledgments}
This work was supported by the Summer Research Internship Program, financed by Erasmus+ and the Mahidol-Bremen Medical Informatics Research Unit (MIRU).



\appendix


\end{document}